\documentstyle[12pt]{article}
\newcommand{\bt}{\begin{tabular}}
\newcommand{\et}{\end{tabular}}
\newcommand{\ba}{\begin{array}}
\newcommand{\ea}{\end{array}}
\newcommand{\be}{\begin{equation}}
\newcommand{\ee}{\end{equation}}
\newcommand{\ben}{\begin{enumerate}}
\newcommand{\een}{\end{enumerate}}

\newtheorem{Thm}{Theorem}[section]

\newcommand{\End}{\mbox{\rm End}\, }
\newcommand{\ad}{\mbox{\rm ad}\, }
\newcommand{\Ldwa}{\,\stackrel{2}{\bigwedge}}
\newcommand{\Ltrzy}{\,\stackrel{3}{\bigwedge}}
\newcommand{\w}{{\!}\wedge{\!}}
\newcommand{\h}{\hspace*{.1mm}}
\newcommand{\hs}{\hspace*{1mm}}

\font \msb=msbm10 scaled \magstep1

\newcommand{\rtimes}{\mbox{\msb o}\,}
\newcommand{\bR}{\mbox{\msb R} }
\newcommand{\bC}{\mbox{\msb C} }

\font \eul=eufm10 scaled \magstep2

\newcommand{\gotG}{\mbox{\eul g}}
\newcommand{\gotH}{\mbox{\eul h}}

\newcommand{\ar}{\alpha }
\newcommand{\tar}{\widetilde{\alpha}}
\newcommand{\br}{\beta }
\newcommand{\gr}{\gamma }
\newcommand{\dr}{\delta }
\newcommand{\er}{\varepsilon }
\newcommand{\lr}{\lambda }
\newcommand{\Lr}{\Lambda }
\newcommand{\om}{\omega }
\newcommand{\Om}{\Omega }

\begin{document}

\title{\bf Poisson Poincar\'{e} groups}
\author{{\bf S. Zakrzewski}  \\
\small{Department of Mathematical Methods in Physics,
University of Warsaw} \\ \small{Ho\.{z}a 74, 00-682 Warsaw,
Poland} }

\date{}
\maketitle

\begin{abstract}

We present almost complete list of normal forms of classical
$r$-matrices on the Poincar\'{e} group.

\end{abstract}

\section*{Introduction}
\vspace{-1mm}
Classification of Poisson-Lie structures on a Lie group is
closely related to the classification of quantum deformations of
the group, cf.\cite{cl,repoi}.  Studying Poisson structures has
here some advantages:
\ben
\vspace{-1mm}
  \item  the notion of a Poisson-Lie group \cite{D:ham}
(see also references in \cite{repoi}) is simple, clear and general,
whereas the notion of a quantum deformation is rather difficult,
varies from author to author, and can be different for different
types of groups (for instance, a separate definition for semidirect
products),
\vspace{-2mm}
  \item calculations of Poisson structures are technically much
easier (classical Yang-Baxter equation is quadratic whereas the
quantum one is cubic) and
have often a direct Lie-algebraic meaning,
\vspace{-2mm}
  \item in some cases it is easy to pass from a Poisson-Lie
group to the corresponding quantum group \cite{luki}:
the classical $r$-matrix can be used to
     \ben
\vspace{-2mm}
      \item construct all remaining objects,
\vspace{-.1mm}
      \item denote the deformation (convenient when
       communicating with other people),
     \een
\vspace{-2mm}
  \item it is easier to check whether the Poisson-Lie group is
non-complete than to check if the corresponding quantum
deformation (on the Hopf $^*$-algebra level) can be formulated
on the C$^*$-algebra level \cite{leni}.
  \vspace{-1mm}
  \een
We point out also, that models of quantum physical systems based
on quantum symmetry correspond usually to (simpler) models of
classical physical systems based on Poisson symmetry and some
ideas of non-commutative geometry can be tested already on the
semi-classical level \cite{poi,poican,k-part}.

The aim of this short report is to present main results of our
study \cite{PsPg} of Poisson structures on the Poincar\'{e}
group. Our classification agrees with the classification of
quantum deformations of the Poincar\'{e} group obtained recently
in \cite{qpoi}.
% It would be interesting to study this connection in more
% detail.
In most cases, the quantum $R$-matrix (in the sense of
\cite{F-R-T}) turns out to be just the exponential of our
corresponding classical $r$-matrix.

% For some basic notation or notions we refer to
% our article \cite{poihom} in the same volume, or, e.g. to
% \cite{repoi,abel,luki}.

For some basic notation or notions we refer to
\cite{poihom,repoi,abel,luki}.

\section{Inhomogeneous $o(p,q)$ algebras}

We consider a $(p+q)$-dimensional real vector space $V\cong \bR
^{p+q}$, equipped with a scalar product $\eta$ of signature
$(p,q)$. Let $\gotH := o(p,q)$ denote the  Lie algebra of the
group $H\cong O(p,q)$ of endomorphisms of $V$ preserving $\eta$,
and let $\gotG := V\rtimes \gotH$ be the corresponding
`inhomogeneous' Lie algebra.
\vspace{-3mm}
\begin{Thm}
(cf.\cite{PsPg}) \
For $\dim V >2$ any cocycle $\dr\colon \gotG\to \Ldwa
\gotG$ is a coboundary:
$$\dr (X) = \ad _X r\qquad \mbox{for}\;\; X\in \gotG.$$
Additionally, for $\dim V > 3$, $r\mapsto  \ad \;r $ is injective.
\end{Thm}
In view of this theorem, the classification of   Poisson
structures on $G=V\rtimes H$ consists in a description of
equivalence classes (modulo ${\rm Aut}\, \gotG$) of $r\in \Ldwa
\gotG$ such that $[r,r]\in (\Ltrzy \gotG )_{\rm inv}$. Here the
subscript `inv' refers to the subset of invariant elements.

We have a decomposition
$$ r = a + b + c ,$$
corresponding to the  decomposition
$$ \Ldwa \gotG = \Ldwa V \oplus (V\wedge \gotH ) \oplus \Ldwa
\gotH .$$
We have also the following decomposition of the Schouten bracket
$$ [r,r] = 2[a,b] + (2[a,c] +[b,b]) +2[b,c] +[c,c],$$
corresponding to the decomposition
$$ \Ltrzy \gotG = \Ltrzy V \oplus (\Ldwa V\wedge \gotH ) \oplus
(V\wedge \Ldwa\gotH ) \oplus \Ltrzy \gotH .$$
Note that
\vspace{-1mm}
$$ (\Ltrzy \gotG )_{\rm inv}= (\Ltrzy V )_{\rm inv}\oplus
(\Ldwa V\wedge \gotH )_{\rm inv} \oplus
(V\wedge \Ldwa\gotH )_{\rm inv} \oplus (\Ltrzy \gotH )_{\rm
inv}.$$
\begin{Thm} (cf.\cite{PsPg})
 If $\dim V > 3$ then  $(\Ltrzy \gotG )_{\rm inv}=
(\Ldwa V\wedge \gotH )_{\rm inv} = \bR \cdot \Om $, where $\Om $
is the canonical element of $ \Ldwa V\wedge \gotH \equiv  \Ldwa
V\otimes \gotH$.
\end{Thm}
We recall that $\Ldwa V$ is naturally isomorphic to $\gotH$ as
a $\gotH$-module. The isomorphism is given by
\vspace{-1mm}
$$ \Ldwa V\ni x\wedge y \mapsto \Om _{x,y}:= x\otimes \eta (y) -
y\otimes \eta (x)\in \End V$$
(here $\eta $ is interpreted as a map from $V$ to $V^*$).
This isomorphism defines a canonical element in $(\Ldwa V)^*\otimes
\gotH $, and, using the identification of $V$ and $V^*$, a
canonical element $\Om \in \Ldwa V \otimes \gotH$. If
$e_1,\ldots ,e_{p+q}$
denotes a basis of $V$, the canonical element $\Om $ is given by
\vspace{-1mm}
$$ \Om = \eta ^{km} \eta ^{ln} e_k\wedge e_l \otimes \Om _{m,n} $$
\vspace{-1mm}
(summation convention), where $\Om _{m,n} := \Om _{e_m ,e_n}$
and $\eta ^{km}$ is the contravariant metric.

{}From the above theorem it follows that Poisson structures on
$G=V\rtimes H$ are in one-to-one correspondence with
$r=a+b+c\in \Ldwa \gotG$ such that
\begin{eqnarray}
  {} [c,c] & = & 0 \label{cc} \\
  {} [b,c] & = & 0 \label{bc} \\
  {} 2[a,c] + [b,b] & = & t\h \Om \qquad (t\in \bR ) \label{bb} \\
  {} [a,b] & = & 0 . \label{ab}
\end{eqnarray}
Equation (\ref{cc}) means that $c$ is a {\em triangular}
$r$-matrix on $\gotH$ (this is the semi-classical counterpart of
a known theorem \cite{inho} excluding the case when the
homogeneous part $H$ is $q$-deformed). Equation (\ref{bc}) tells
that $b$, as a map from $\gotH ^*$ to $V$, is a cocycle, the Lie
bracket on $\gotH ^*$ being defined by the triangular $c\in \Ldwa
\gotH$ and the action of $\gotH ^*$ on $V$ is defined using the
homomorphism from $\gotH ^*$ to $\gotH$ given by $c$.

Let us list some particular cases.
\ben
\vspace{-1mm}
 \item $b=0$, $c=0$, $a\in \Ldwa V$ arbitrary. This type of
solutions we call `soft deformations' \cite{abel}.
 \item $a=0$, $c=0$, $[b,b]=t\Om $. There is a family of
solutions of the latter equation, parametrized by vectors in
$V$. Namely, for each $x\in V$,
\be\label{bx}
 b_x:= \eta ^{kl} e_k\otimes \Om _{e_l,x}
\ee
($e_k$ is any basis in $V$) satisfies this equation with $t=
-\eta (x,x)$.
\item $a=0$, $b=0$, $c\in\Ldwa \gotH$ triangular.
\een

\section{The case of the Poincar\'{e} group}

Using the list of classical $r$-matrices for the Lorentz group
from \cite{repoi}
(only triangular are needed), we have solved equations
(\ref{cc})--(\ref{ab}) in the case of the Poincar\'{e} group,
assuming $c\neq 0$ or $t=0$, and we have found several solutions
in the case $t\neq 0$.
The results are shown in the table below.
Examples with $t\neq 0$ are
provided by formula (\ref{bx}). Let $e_0$, $e_1$, $e_2$, $e_3$
be a Lorentz basis in $V$. Let us introduce the standard
generators of $\gotH$:
$$M_i = \er _{ijk} e_k\otimes e{^j},\qquad  L_i = e_0\otimes e{^i} +
e_i\otimes e{^0}$$
($i,j,k=1,2,3$).
 If we set $x:= e_0$ in (\ref{bx}), we obtain
\vspace{-1mm}
 $$ b_{e_0}  = \sum_{k=1}^3 L_k \wedge e_k ,$$
 \vspace{-1mm}
 which is the known \cite{luki} classical $r$-matrix
corresponding to so called $\kappa$-deformation. Taking $x=e_3$,
we obtain another solution
$$ b_{e_3} = M_1\wedge e_2 - M_2\wedge e_1 + L_3 \wedge e_0$$
(this one is $L_1,L_2,M_3$-invariant). As shown in the table
below, both $b_{e_0}$ and $b_{e_3}$ are particular cases of more
general families (thus, we have a `deformation' of the
$\kappa$-deformation).

The following table lists 23 cases labelled by the number {\sf
N} in the last column. The question mark in the table (case 9)
reminds that the case $c=0$, $[b,b]\neq 0$ is not yet completely
solved (including the question mark, the list is complete).

\vspace{5mm}

\noindent
{\footnotesize
%\bt{|r|c|c|c|c|c|c|}
\bt{|@{\h}c@{\h}|@{\h}c@{\h}|@{\h}c@{\h}|@{\h}c@{\h}|@{\h}c@{\h}|@{\h}c@{\h}|}
  \hline
%  No. &
 $c$ & \multicolumn{2}{c|}{$b$} & $a$  & \#
 & {\sf N} \\
  \hline\hline
%  1 &
 $\gr JH\w H$ & \multicolumn{2}{c|}{$0$} &
     $\ar e_+\w e_- + \tar e_1\w e_2 $ & 2
 & 1      \\
  \hline
%  2 &
 $JX_+\w X_+ $ & \multicolumn{2}{c|}{$\br_1 (e_1\w
     X_+ - e_2\w JX_+ +  e_+\w H) + \br _2 e_+\w
   JH $} & 0 &  1 & 2 \\
   \cline{2-6}
%  3 &
 & \multicolumn{2}{c|}{$\br (e_1\w X_+ - e_2\w
 JX_+ +e_+\w H)$} & $\ar e_+ \w e_1 $ & 1 & 3 \\
   \cline{2-6}
%  4 &
 & \multicolumn{2}{c|}{$\br (e_1\w X_+ + e_2\w
 JX_+)$} & $e_+ \w (\ar _1 e_1 + \ar _2 e_2)- \br ^2
e_1\w e_2 $ &  2  & 4 \\
 \hline
  $H\w X_+ - $  &
 \multicolumn{2}{c|}{} &  &    &  \\
% 5 &
 $ JH\w JX_+ +  $  &
 \multicolumn{2}{c|}{$0$} & $0$ &  1  & 5 \\
  $ \gr JX_+\w X_+ $  &
 \multicolumn{2}{c|}{} & &   &  \\
 \hline
%  6 &
 $H\w X_+$ & \multicolumn{2}{c|}{$\br e_2\w X_+$} &
0 & 0  & 6 \\
 \hline
% 7 &
 $0$ &\hs $t\neq 0$\hs & $e_1\w L_1+e_2\w L_2 + e_3\w
 L_3 +\br e_0\w M_3 $ & 0 & 1  & 7 \\
   \cline{3-6}
% 8 &
 &  & $e_2\w M_1 -e_1\w M_2 + e_0\w
 L_3 +\br e_3\w M_3 $ & 0 & 1  & 8 \\
    \cline{3-6}
% 9 &
 &  & {\bf ?} & ? &  ?  & 9 \\
    \cline{2-6}
% 10 &
 & \hs $ t = 0$\hs & $e_1\w X_+ - e_2\w JX_+ + e_+\w
(H+\br JH)$ & $0$ & 1  & 10 \\
   \cline{3-6}
% 11 &
  & & $ e_1\w (X_+ +\br _1 JX_+) + e_+\w (H+\br
_2X_+)$ & $ \ar e_+\w e_2$ &  2  & 11 \\
 &  & $\br _2=0$ or $\br _2=\br_1$ or $\br_2=\pm 1$ & &  &  \\
    \cline{3-6}
% 12 &
 & & $ e_1\w JX_+ + e_+\w X_+$ & $ e_-\w (\ar _1
e_1 + \ar _2e_2 )-\ar _2e_+\w e_- +\ar e_+\w e_2$ & 3
 & 12 \\
   \cline{3-6}
% 13 &
 & & $ e_0\w JH$ & $ \ar_1e_0\w e_3 + \ar _2
e_1\w e_3 +\ar _3 e_1\w e_2$ &  3  & 13 \\
   \cline{3-6}
% 14 &
  & & $ e_1\w H$ & $ \ar_1e_0\w e_3 + \ar _2
e_1\w e_2 +\ar \om $ &  3  & 14 \\
 & & &   $\om =e_0\w e_2$, $\om = e_2\w e_3$, $\om =
e_2\w e_+$ &  &  \\
 \cline{3-6}
% 15 &
 & & $ e_2\w X_+$ & $ \tar e_+\w e_- +\ar e_+\w
e_1 + e_-\w (\ar_1e_1 + \ar _2e_2)$ &  2
 & 15 \\
 \cline{3-6}
% 16 &
 & & $ e_3\w JH$ & $ \ar_1e_0\w e_3 + \ar _2
e_0\w e_1 +\ar _3 e_1\w e_2 $ &  3  & 16 \\
 \cline{3-6}
% 17 &
 & & $ e_+\w X_+$ & $e_-\w (\ar_1e_1 +\ar_2e_2)
 +\ar e_+\w e_2$ & 2 & 17 \\
  \cline{4-6}
% 18 &
 & & & $\ar _1 e_1\w e_2 +\ar_2 e_+\w e_2
 +\ar _3 e_+\w e_2$ & 2  & 18 \\
 \cline{3-6}
% 19 &
 & & $ e_+\w (\br _1H+\br _2JH)$ &
  $\ar e_1\w e_2 + \ar _1 e_-\w e_1$ & 2
  & 19 \\
    \cline{3-6}
% 20 &
 & & $ e_+\w JH$  &
  $\ar e_1\w e_2 + \ar _1 e_-\w e_1
  +\tar e_+\w e_- $ &  2 & 20 \\
      \cline{3-6}
% 21 &
 & & $ e_+\w H$ &
  $\ar e_1\w e_2 + \ar _1 e_-\w e_1
  + e_+\w (\tar _1e_1 + \tar _2e_2) $ & 3 & 21 \\
      \cline{2-6}
% 22 &
 & \multicolumn{2}{c|}{$0$} & $\ar _1 e_0\w e_3 + \ar
_2 e_1\w e_2$ &  1 & 22 \\
      \cline{4-6}
% 23 &
 & \multicolumn{2}{c|}{} & $e_1\w e_+$ & 0 & 23 \\
 \hline
\et
}

\vspace{5mm}

Now we explain the notation in the table. We have introduced the
standard generators of $\gotH = sl(2,\bC )$:
\vspace{-1mm}
$$H=\frac12 \left[ \ba {cc} 1 & 0 \\ 0 & -1 \ea \right] ,\qquad
X_+=\left[ \ba {cc} 0 & 1 \\ 0 & 0 \ea \right] ,\qquad
X_-=\left[ \ba {cc} 0 & 0 \\ 1 & 0 \ea \right] $$
(recall that the action of $X\in sl(2,\bC )$ on a vector $v\in
V$ is given by $X(v):= Xv + vX^+$, the space $V$ being
identified with the set of hermitian $2\times 2$ matrices, where
$X^+$ is the hermitian conjugate of $X$).
We denote by $J$ the multiplication by the imaginary unit in
$\gotH$. It is also convenient to introduce the light-cone
vectors $e_{\pm } := e_0 \pm e_3$.

In the forth column (labelled by {\#}) we indicate the number of
essential parameters (more precisely -- the maximal number of
such parameters)
involved in the deformation. This number is in many cases less
than the number of parameters actually occurring in the table.
The reduction of the number of parameters can be achieved using
two following one-parameter groups of automorphisms of $\gotG$:
\begin{itemize}
 \item the group of dilations: $(v,X)\mapsto (\lr v,X)$
   \ (in cases 1,2,3,4,6,17,18,19,20,21,22),
 \item the group of internal automorphisms generated by $H$ \ (in
cases 15,17,18,19,20 21).
\end{itemize}

In order to make some comments on the relations between our
classical $r$-matrices and the quantum Poincar\'{e} groups, let
us set  $R:=\exp (ir)\in \End (\bC ^5\otimes \bC ^5)$. We have
two following remarks.
\ben
 \item If $c=0$ (cases 7--23), then $r^2=0$, hence $R=1+ir$ and
the corresponding to this $R$-matrix
commutation relations for the elements of the $5\times 5$ matrix
$$ T = (T^a{_b})_{a,b=0,\ldots ,4} = \left(\ba{cc} \Lr & v \\ 0
& 1 \ea\right) ,\qquad \Lr = (\Lr ^\mu _{\;\nu})_{\mu ,\nu
=0,\ldots ,3},\qquad v = (v^{\mu})_{\mu =0,\ldots ,3}  $$
arise simply by replacing the Poisson brackets (defined by $r$)
by commutators (divided by $\sqrt{-1}$). As in \cite{luki}, no
ordering ambiguities arise in this case. Moreover, the
right-hand-sides of the expressions for commutators satisfy
(automatically) the Jacobi identity (the computation is the same
as in the Poisson case in which it is true because we started
with a Poisson structure). This is sufficient to show that the
resulting algebra has a `correct size'.

Furthermore, from $r_{12}r_{13}r_{23}=0=r_{23}r_{13}r_{12}$ it
follows that
$$R_{12}R_{13}R_{23}=1+r_{12} + r_{13} + r_{23} + r_{12}r_{13} +
r_{12}r_{23} + r_{13}r_{23},$$
$$R_{23}R_{13}R_{12}=1+r_{23} + r_{13} + r_{12} + r_{23}r_{13} +
r_{23}r_{12} + r_{13}r_{12},$$
and
$$ R_{12}R_{13}R_{23} - R_{23}R_{13}R_{12} = [r,r].$$
Hence $R$ satisfies the Yang-Baxter equation if and only if
$[r,r]=0$ (cases 10--23).
\item If $c\neq 0$, then we can compare our results with
\cite{qpoi} (in \cite{qpoi}, only quantum Poincar\'{e} groups
with non-trivial `Lorentz part' were classified). Using
\cite{repoi} we have a clear correspondence between four types
of triangular classical $r$-matrices $c$ on the Lorentz Lie
algebra and four types of triangular ($q=1$) quantum Lorentz
groups \cite{cl}.  For a fixed quantum Lorentz group
(corresponding to our $c$), the classification of quantum
Poincar\'{e} groups given in \cite{qpoi} involves quantities
similar to our $b$ and $a$ (they are denoted by $H$ and $T$) and
one can observe a clear correspondence between particular
solutions presented in \cite{qpoi} and the cases 1--6 of our
table. Whether and how the quantum $R$-matrix can be constructed
from $r$ is not completely clear yet. One can check that in the
first case, $R:=\exp (ir)$ coincides with the $R$-matrix
obtained in \cite{qpoi}. The same seems to be true for the
second type of $c$. In two remaining cases of $c$ the relation
may be more complicated, because already the $R$-matrix for the
Lorentz part differs from $\exp (ic)$ (however it is built of
components $\exp (ic_-)$, $\exp (ic_+)$, see \cite{repoi}).
\een

\end{document}